\documentclass[aps,prb,twocolumn,superscriptaddress]{revtex4-1}

\usepackage{hyperref} % hyperreferences
\usepackage{graphicx}
\usepackage{amsmath}
\usepackage{amssymb}

\DeclareGraphicsExtensions{.png,.jpg,.eps}
\usepackage{xcolor}

\begin{document}

\author{J. L. Lado}
\affiliation{Department of Applied Physics, Aalto University, Espoo, Finland}

\author{M. Sigrist}
\affiliation{Institute for Theoretical Physics, ETH Zurich, 8093 Zurich, Switzerland}

\title{
	Solitonic in-gap modes in a superconductor-quantum
	antiferromagnet interface
}

\begin{abstract}
        Bound states at interfaces between superconductors and other materials are a powerful tool
        to characterize the nature of the involved systems, and to engineer elusive quantum
	excitations.
        In-gap excitations of conventional s-wave superconductors occur, for instance,
        at magnetic impurities with net magnetic moment
        breaking time-reversal symmetry.
        Here we show that interfaces between a superconductor
        and a quantum antiferromagnet can host robust
        in-gap excitations, without breaking
        time-reversal symmetry.
        We illustrate this phenomenon in a one-dimensional model
        system with an interface
        between a conventional s-wave superconductor
        and a one-dimensional Mott insulator described by a standard
        Hubbard model. This genuine many-body problem is solved exactly by
	employing a combination
        of kernel polynomial and tensor network techniques.
        We unveil the nature
        of such zero modes by showing that they
        can be adiabatically connected
        to solitonic solutions between a superconductor
        and a mean-field antiferromagnet. Our results put forward
        a new class of in-gap excitations between
        superconductors and a disordered quantum spin phase,
	including quantum spin-liquids, that
        can be relevant for a wider
        range of heterostructures.

 \end{abstract}

\date{\today}

\maketitle

%%%%%%%%%%%%%%%%%%%%%%%%%%%%%%%%%%%%%%%%%%%%%%%%
%%%%%%%%%%%%%%%%%%%%%%%%%%%%%%%%%%%%%%%%%%%%%%%%
%%%%%%%%%%%%%%%%%%%%%%%%%%%%%%%%%%%%%%%%%%%%%%%%

\section{Introduction}

Topological modes emerging in condensed matter systems are among the
most intriguing features in physics. Well-known examples are the electronic solitons in
polyacetylene\cite{PhysRevLett.42.1698} 
or the Jackiw-Rebbi modes first
introduced in high-energy theory\cite{PhysRevD.13.3398}.
In recent years the family of topological phases with extraordinary modes has been extended
enormously to a multitude of novel systems with gapped bulk excitation spectra \cite{RevModPhys.83.1057,RevModPhys.82.3045,Ando2015}. 
In almost all cases, such topological modes emerge in systems that can be described within the spectra 
of non-interacting electrons, whose single-particle Hamiltonians incorporate a non-trivial topology. 
Despite the large body of knowledge on topologically non-trivial excitations of non-interacting
particles accumulated in recent years, the theoretical analysis of the many-body counterpart 
remains a formidable challenge.  

Among the different in-gap states found in materials, those of superconductors have attracted special
attention, because they might provide valuable information about the nature of the superconducting phase,
even if it is topologically trivial. On the one hand, 
a classical magnetic impurity (a static magnetic moment) gives rise to
in-gap Yu-Shiba-Rusinov states in s-wave superconductor, probing the vulnerability to the superconducting 
phase against time-reversal symmetry violation (spin polarization)\cite{PhysRevLett.118.117001,
PhysRevLett.120.167001,PhysRevLett.115.087001,
PhysRevLett.110.217005,PhysRevLett.117.186801,PhysRevLett.120.156803,PhysRevLett.107.256802,PhysRevLett.119.197002}.
On the other hand, in-gap states created by non-magnetic impurities provide a strong signature
for unconventional superconductivity \cite{1965JETP130L,PhysRevB.37.4975,PhysRevB.48.653,PhysRevLett.80.161}

Increasing complexity, for instance, through heterostructures connecting a superconductor to 
materials of various properties offers an attractive platform
to create new emergent phases \cite{PhysRevLett.105.077001,NadjPerge2014,PhysRevLett.100.096407}.
This is the basis for a plethora of proposals to engineer Majorana bound states \cite{Alicea2012}, to explore unusual Andreev
physics\cite{PhysRevLett.105.097002,PhysRevLett.109.087004,PhysRevLett.109.106404,PhysRevLett.105.097002,PhysRevB.70.012507} and even
to design higher-dimensional topological superconductors.\cite{Schindler2018,PhysRevB.100.205406}
So far studies in this direction have focused mainly on single-particle physics,
\cite{RevModPhys.83.1057,RevModPhys.82.3045,Ando2015}
e.g. system in which the excitation spectrum can be treated in a mean-field picture. 
Therefore, extending the scope to interface physics involving the strongly correlated electron 
regime with dominant quantum fluctuations represents a rich playground for new physics
which is largely unexplored \cite{PhysRevLett.114.066401,PhysRevB.88.161103,PhysRevB.89.115430,PhysRevB.75.024505,PhysRevB.89.220504}.

\begin{figure}[t!]
\centering
    \includegraphics[width=\columnwidth]{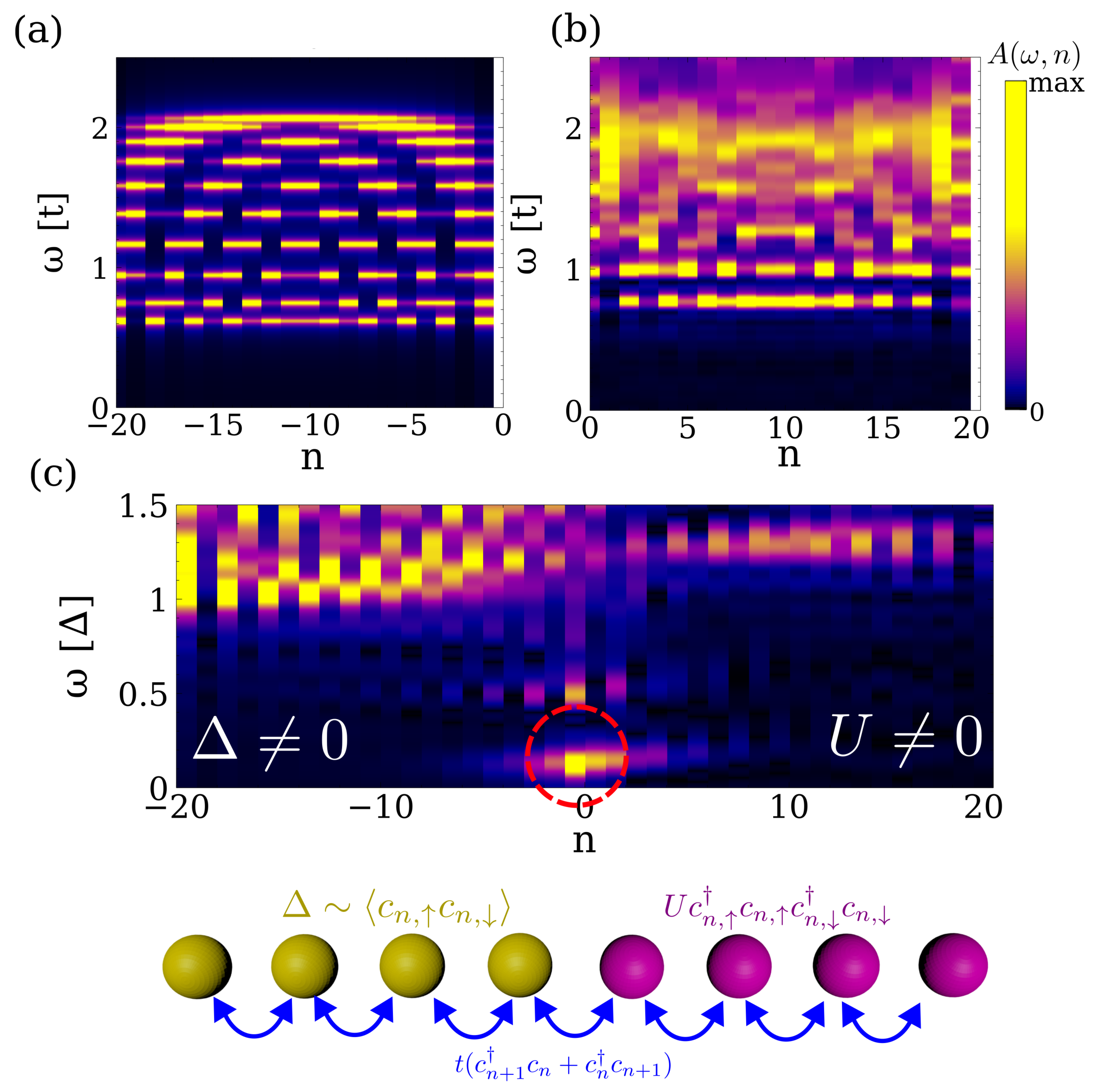}

\caption{
	(a) Spectral function of the superconductor,
	showing a gap up to the superconducting gap $\Delta$,
	and of the quantum antiferromagnet (b), showing
	a gap on the order of the Hubbard $U$.
	Panel (c) shows the spectral function across the interface
	between the two systems, showing the 
	emergence of in-gap modes at the interface
	in the absence of time reversal symmetry breaking.
	We use $\Delta=0.6t$ and $U=4t$.
	The inset below panel (c) shows a sketch of the model
	used to study the superconducting-antiferromagnetic
	interface.
}
\label{fig1}
\end{figure}

Here we demonstrate how solitonic in-gap modes can emerge at interfaces between
a conventional superconductor and a quantum antiferromagnet without long-range order, both topologically trivial on their own. 
In particular, we show that time-reversal symmetry needs not to be broken and that these modes
can be adiabatically connected with solitonic zero modes of the 
antiferromagnetically ordered phase violating time-reversal symmetry. In this way, we extend the set
of situations where the composition of different materials can generate a non-trivial phase at interfaces. 
In particular, our results put forward a minimal model system where the
interplay of superconductivity and quantum spin lqiuid physics gives
rise to unconventional excitations.

The manuscript is organized as follows. In Sec. \ref{sec:exact} we show
the emergence of the solitonic zero mode and its robustness
towards system parameters, by exactly solving the
interacting model.
In Sec. \ref{sec:ori} we put forward a connection between the 
time-reversal symmetric interacting zero
mode, and a solitonic zero mode 
in a non-interacting model with broken
time-reversal symmetry. Finally, in Sec. \ref{sec:con}
we summarize our conclusions.

\section{Emergence of a solitonic mode}
\label{sec:exact}

We model our system by the following Hamiltonian of a one-dimensional chain, that allows us 
incorporate an interface between a conventional superconductor and 
a quantum antiferromagnet in the simplest way: $ H = H_{kin} + H_U + H_{SC} $,
where $H_{kin}$ is the kinetic energy term in a tight-binding form,
\begin{equation}
H_{kin} = t\sum_ {n,s} [c^\dagger_{n,s}c_{n+1,s} +
c^\dagger_{n+1,s}c_{n,s}] + 
\sum_{n,s} \mu(n)
c^\dagger_{n,s}c_{n,s} \, ,
\label{hkin}
\end{equation}
$H_{U}$ is the Hubbard interaction term with a position dependent $ U $
\begin{equation}
H_{U} = 
\sum_{n,s} U(n)
c^\dagger_{n,\uparrow}c_{n,\uparrow}
c^\dagger_{n,\downarrow}c_{n,\downarrow} \, ,
\label{hu}
\end{equation}
and $H_{SC}$ introduces conventional superconductivity in the mean-field formulation
\begin{equation}
H_{SC} = 
	\sum_n\Delta(n) [c_{n\uparrow} c_{n\downarrow} +
c^\dagger_{n\downarrow} c^\dagger_{n\uparrow} ] \, .
\label{hsc}
\end{equation}
The heterostructure
can be modeled by the parametrization
$U(n) = [\tanh(n/W) + 1]U/2$,
and $\Delta(n) = [-\tanh(n/W) +1]\Delta/2$ locating the interface at $n= 0 $
(Fig.\ref{fig1}(a)),
and we take $W=1$. The profile of $\mu(n)$ is chosen
as $\mu(n) = -U(n)/2$
so that the system is
half filled everywhere.
Our calculations are performed in chains having 40 sites.

For the treatment of this genuine many-body Hamiltonian 
we employ the computational matrix product state formalism and, in particular, 
determine the local single-particle spectral function defined as
\begin{equation}
	A(\omega,n) = 
	\sum_s
	\langle GS | c^\dagger_{n,s} \delta(\omega - H+E_{GS})
 c_{n,s} |GS \rangle \; .
	\label{eq:a}
\end{equation}

This dynamical correlation function can be computed for the
whole frequency range by exploiting a kernel polynomial
technique \cite{RevModPhys.78.275} implemented within the matrix product
state formalism of ITensor \cite{ITensor,dmrgpy}.
The basic idea of the method consists of representing
the function $A(n,\omega)$ in a complete
functional basis expanded by $N$ Chebyshev polynomials
$T_k(\omega)$ as 
$
A(\omega,n)=
        \frac{1}{\pi\sqrt{1-\omega^{2}}}
        \left(\mu_{0}+2\sum_{k=1}^{N}\mu_{k}T_{k}(\omega)\right)
$. The coefficients $\mu_k$ are obtained
as
$\mu_{k}=\langle GS| c^\dagger_n T_{k}(H) c_n| GS \rangle$,
\footnote{The Hamiltonian must be scaled to the interval (-1,1)
to perform the Chebyshev expansion.}
that can be recursively computed through
products of matrix product operators
and matrix product states \cite{RevModPhys.78.275,PhysRevB.90.115124,PhysRevB.91.115144,PhysRevResearch.1.033009} .
Note that for this algorithm the time evolution is not needed, since we work 
from the beginning in frequency space.

The spectral function $A(n,\omega)$ shows the quasiparticle excitation gap in real space. 
Thus, it is instructive to consider first each subsystem of the model
separately using our computational scheme for a system of finite length. For
the
uniform superconductor, $A(\omega,n) $ shows a quasiparticle gap (Fig.\ref{fig1}(a)).
The half-filled Hubbard chain with $ U >0$ is not magnetically ordered, but
displays a Mott charge excitation gap as seen in
Fig.\ref{fig1}(b)\cite{PhysRevLett.77.1390,PhysRevLett.120.046401,PhysRevLett.72.2765}.
Note that the spatial dependence of the spectral functions in Fig.\ref{fig1}(a,b) is a finite size effect induced by
the open boundary conditions. It is also interesting to note that
both systems are topologically trivial, lacking in-gap edge modes.

\begin{figure}[t!]
\centering
    \includegraphics[width=\columnwidth]{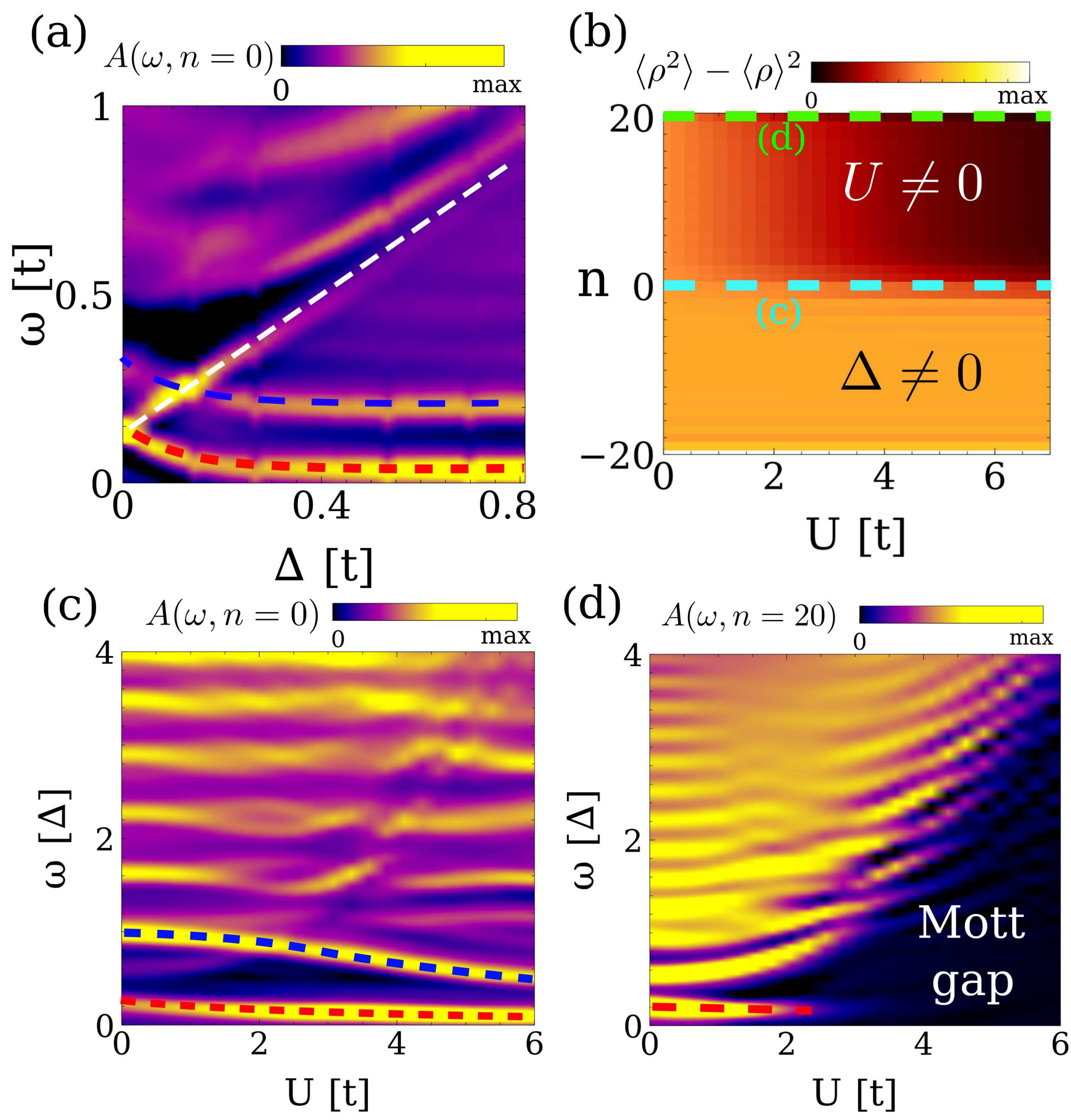}

\caption{
	(a) Evolution of the spectral function
	at the interface as a function of
	the superconducting pairing $\Delta$,
	for the other half chain with finite $U$.
	Panel (b,c,d) show the evolution as a function of $U$
	with the superconducting part at a fixed finite $\Delta$.
	Panel (b) shows the spatially resolved charge fluctuation in (b),
	the spectral function at the interface in (c) and
	the spectral function in the quantum antiferromagnetic part
	in (d).
	We use $U=6t$ for (a) and
	$\Delta=0.4t$ for (b,c,d).
}
\label{fig2}
\end{figure}

We turn now to the spatially resolved spectrum of a heterostructure
connecting the two phases. As shown in Fig.\ref{fig1}(c),
the system shows now in-gap excitations (the lowest one
highlighted with the
dashed red circle), 
which are clearly a feature connected with the interface ($n \approx 0 $). 
Besides the previous in-gap mode, 
a second in-gap state at a higher energy can be observed
at the interface in Fig.\ref{fig1}(c).
In-gap states in an s-wave superconductor are usually attributed to static
magnetic impurities, giving rise to the so-called Yu-Shiba-Rusinov states.
In our case, however, time-reversal symmetry is conserved and there
are no static moments despite the suppression of charge fluctuation 
on the Mott side ($n>0$). Moreover, the dominant mode here is essentially  
pinned at zero, a feature that does not happen
for generic Yu-Shiba-Rusinov states.

\begin{figure}[t!]
\centering
    \includegraphics[width=\columnwidth]{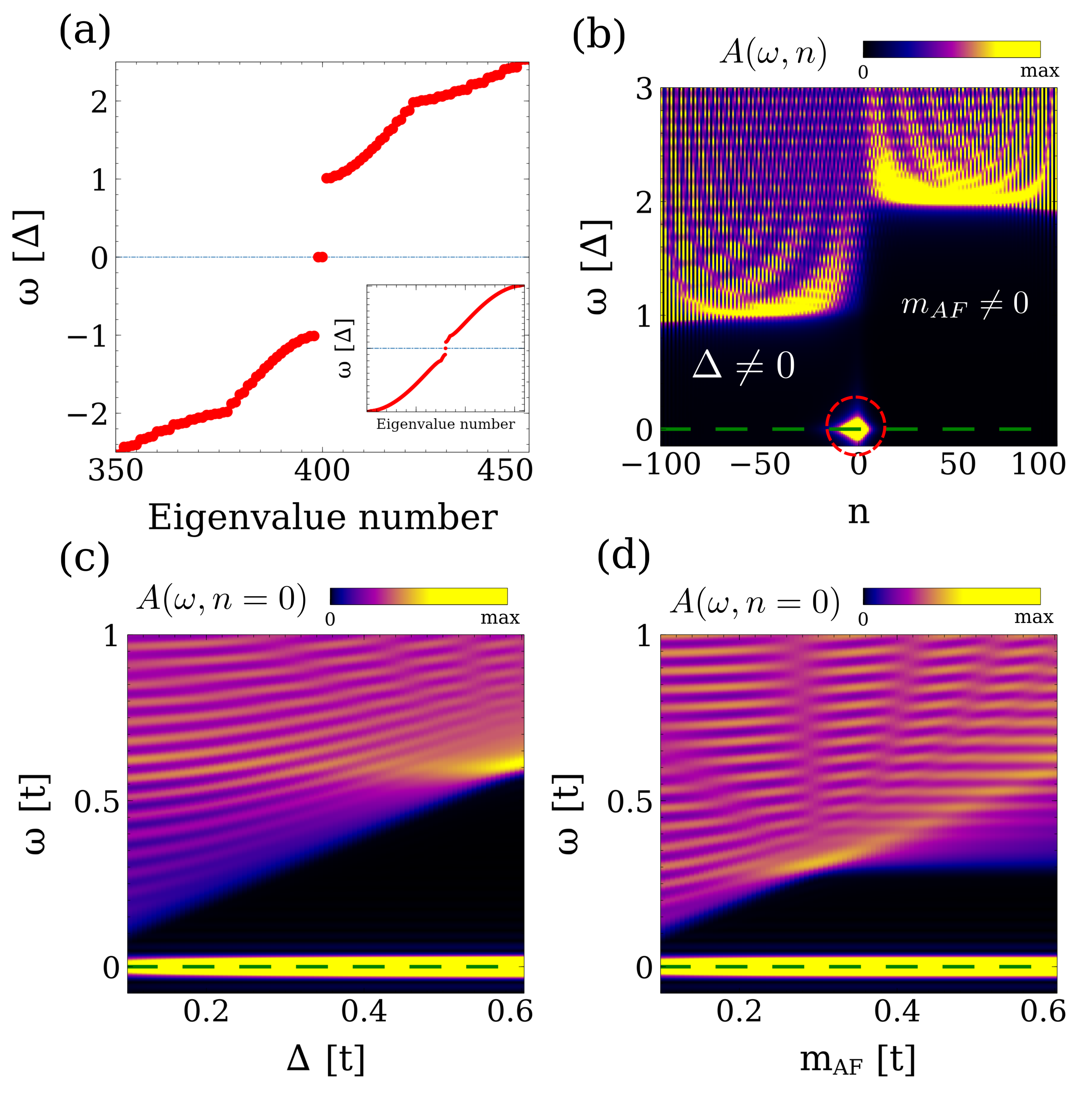}

\caption{
	(a) Bogoliubov de Gennes spectra of
	the superconductor-stagger antiferromagnet
	interface, showing the existence of a zero mode.
	Panel (b) shows the spatially resolved
	density of states $A(\omega,n)$ of the interface,
	showing that the zero mode is
	localized at the interface between the two systems.
	Panel (c,d) show the evolution of the
	density of states at the interface
	as a function of the superconducting pairing $\Delta$
	(c) and the antiferromagnetic stagger field $m_{AF}$
	(d), highlighting the robustness of the zero mode.
	We use for (a,b) $m_{AF}=0.4t$ and $\Delta=0.2t$,
	for (c) $m_{AF}=0.8t$ and for (d) $\Delta=0.3t$.
	Note that the $\omega$ axis starts slightly below
	$\omega=0$ for
	visibility, dark green dashed lines mark $\omega=0$
	in (b,c,d).
}
\label{fig3}
\end{figure}

Let us now consider the properties of this interface excitation. First, we examine
how the interface mode behaves for varying the model parameters $ \Delta $ and $U$. 
In Fig.\ref{fig2}(a) the spectral function $A(\omega,n=0) $ as
function of $ \Delta $ for fixed $U=6t$ shows the evolution of the lowest mode toward $ \omega = 0 $
upon increasing $ \Delta $. 
With increasing $\Delta$ the 
two in-gap modes (red and blue dashed line in Fig. \ref{fig2}(a))
converge to stable in-gap energies, while the bulk superconducting 
gap increases (white dashed line, note the shift
due to finite size effects).
Figs.\ref{fig2}(b-d) display the $ U $-dependence for fixed $ \Delta = 0.4 t $.
In Fig. \ref{fig2}(b) we show how the charge fluctuations are gradually suppressed in the Mott region, while
they remain constant in the superconducting region. The zero-energy mode (red dashed line) 
also settles at the interface
upon increasing $ U $ as shown in Fig.\ref{fig2}(c),
and similar behavior happens with the next in-gap
state (blue dashed line).\footnote{This second in-gap mode
is sensitive to the boundary conditions at the interface as discussed
in the appendix}.
For comparison, we observe that the low-energy modes progressively
fade away in the interior of the Mott region when $ U $ is increased (see Fig.\ref{fig2}(d)).

\section{Origin of the solitonic mode}
\label{sec:ori}

A further path to elucidate the character of the zero-energy modes runs via the
using a mean-field antiferromagnetic phase for $ n > 0 $. We restrict to the single-particle
description by replacing $ H_{U} $ by 
$
H_{AF} =
  \sum_n
(-1)^n m_{AF}(n)  [
c^\dagger_{n\uparrow} c_{n\uparrow} -
c^\dagger_{n\downarrow} c_{n\downarrow}]
$
with the spatial profile $m_{AF} (n) = [1+\tanh(n/W)]m_{AF}/2$ and
$\Delta (n) = [1-\tanh(n/W)]\Delta/2$.
The Hamiltonian for this inhomogeneous 1D system can be easily solved
numerically by means of a Bogoliubov de Gennes (BdG) scheme
with the results displayed in Fig.\ref{fig3}. We find zero-energy modes
within the gap in the sequence of eigenvalues (Fig.\ref{fig3}(a)) and can locate them
clearly at the interface (Fig.\ref{fig3}(b)). When changing the system parameters 
$ \Delta $ for fixed $ m_{AF} =0.8t $ (Fig. \ref{fig3}(c)) and $ m_{AF} $ 
for fixed $ \Delta = 0.3t $ (Fig.\ref{fig3}(d))
we observed that this mode remains
solidly at $ \omega =0 $, which demonstrates clearly that this feature is not
an effect of fine-tuning. 

The nature of this interface mode can be easily explained with an analytical approach in the continuum limit of this model. 
For this purpose we choose a two-site unit cell adapted to the
staggered moment ($A$ and $B$ sublattice) and rewrite the kinetic energy term in $k$-space,

\begin{equation}
H(k) = 
\begin{pmatrix}
	0 & 1+e^{ik} \\
	1+e^{-ik} & 0 
\end{pmatrix}
\end{equation}
which near $k=\pi$ takes the form of a 1D Dirac equation with
$H(p) = \tau_y p$, with $\tau_y$ the sublattice Pauli matrix. We now use $p=-i\partial_x$,
introduced into the adapted $ H_{AF} $ and $ H_{SC}$ and turn to continuum 
variables 
$c_{2n} \rightarrow \psi_A(x)$,
$c_{2n+1} \rightarrow \psi_B(x)$,
$\Delta(n) \rightarrow \Delta(x)$,
$m_{AF}(n) \rightarrow m_{AF} (x)$
defining the continuum 1D Hamiltonian,

\begin{equation}
\begin{split}
	H = 
	\sum_{s,\alpha,\beta}
	p \tau_y^{\alpha\beta} \psi^\dagger_{\alpha,s} \psi_{\beta,s}
	+
	\sum_{s,\alpha}
	m_{AF}(x) \sigma_z^{ss} \psi^\dagger_{\alpha,s} \psi_{\alpha,s}
	+
	\\
	\sum_\alpha
	\Delta(x) [
	\psi_{\alpha,\uparrow} \psi_{\alpha,\downarrow}
	+
	\psi^\dagger_{\alpha,\downarrow} \psi^\dagger_{\alpha,\uparrow}
	]
\end{split}
\end{equation}

\begin{figure}[t!]
\centering
    \includegraphics[width=\columnwidth]{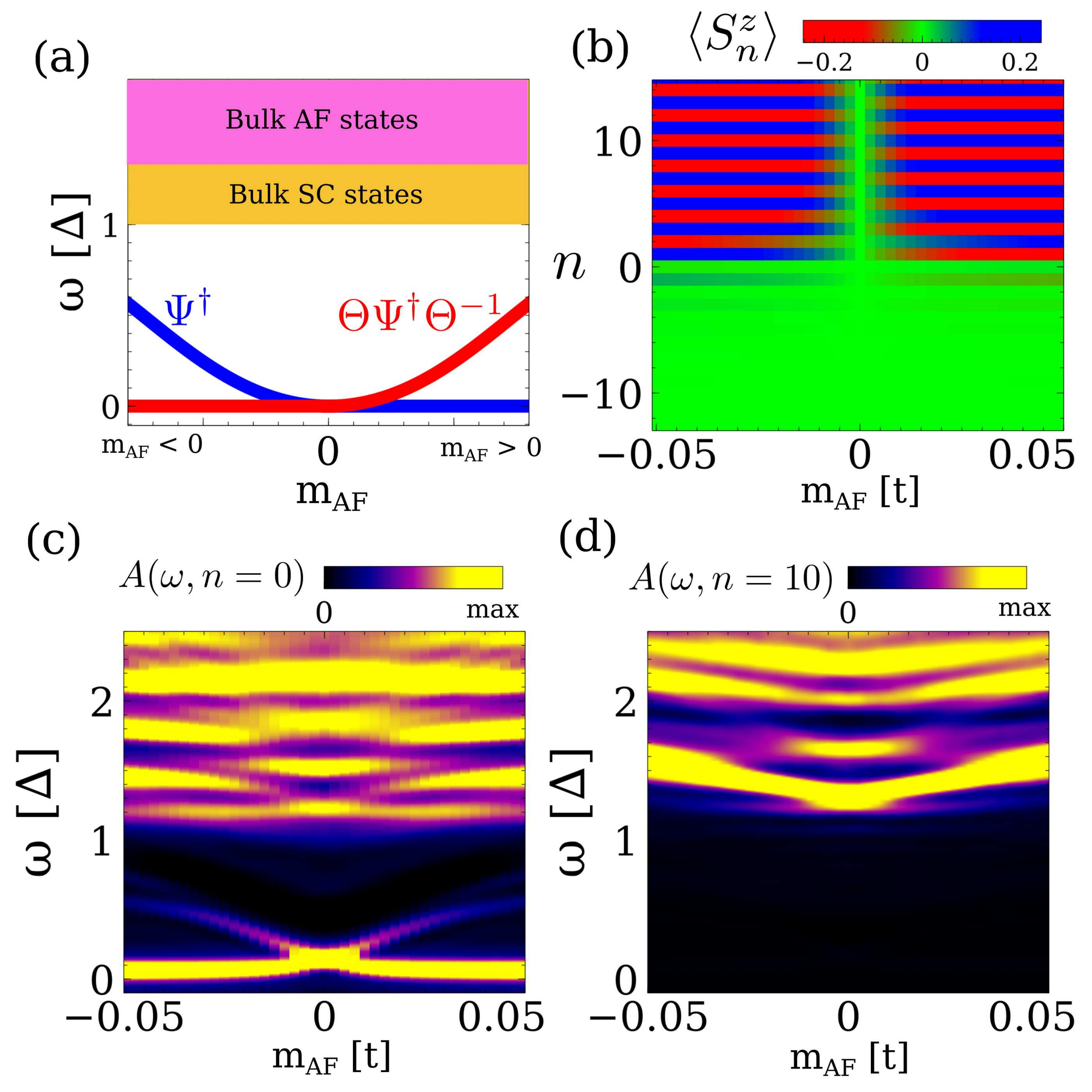}

	\caption{(a) Sketch of the evolution
	of the charge excitation in the heterostructure,
	including the solitonic zero modes,
	as a function of the
	antiferromagnetic field $m_{AF}$,
	showing that they become the in-gap
	excitations in the case of the quantum antiferromagnet.
	(b) Spatially resolved
	magnetization as a function of $m_{AF}$.
	Panels (c,d) show the spectral function
	at the interface (c) and in the middle of the
	antiferromagnetic region (d).
	We use $U=6t$ and $\Delta=0.6t$.
}
\label{fig4}
\end{figure}

This Hamiltonian can be diagonalized defining the Nambu spinor
$\Psi = (
\psi_{A,\uparrow},
\psi_{B,\uparrow},
\psi^\dagger_{A,\downarrow},
\psi^\dagger_{B,\downarrow}
)$,
for the sector of spin-up electron/spin-down hole, where we obtain
$H \sim \Psi^\dagger \mathcal{H} \Psi$ with
\begin{equation}
	\mathcal{H} =
	\begin{pmatrix}
		m_{AF} (x)  & ip & \Delta(x) & 0 \\
		-ip & -m_{AF} (x) & 0 & \Delta(x) \\
		\Delta(x) & 0 & m_{AF} (x)  & -ip \\
		0 & \Delta(x) & ip & -m_{AF} (x)   
	\end{pmatrix} \;.
	\label{solh}
\end{equation}
The spectrum is obtained by BdG transformation. While both the superconductor and the antiferromagnet have
an excitation gap, we find at the interface a zero-energy eigenvalue with an eigenoperator \cite{PhysRevD.13.3398,PhysRevX.5.041042,PhysRevLett.121.037002,PhysRevB.100.125411},
\begin{equation}
	\Psi^\dagger =
\frac{1}{2} [
c^{\dagger}_{A,\uparrow}
+ c^{\dagger}_{B,\uparrow}
-c_{A,\downarrow} +
c_{B,\downarrow} ]
e^{\int_{0} ^x [\Delta (x') - m_{AF} (x')] dx'}
	\label{soliton}
\end{equation}
for $m_{AF} (\infty)>0$.\footnote{For $m_{AF} (\infty)<0$,
the $m_{AF}$ in the exponent gets an additional minus sign to keep
the wavefunction normalized.}
Note that since Eq. \ref{solh} is a real
differential equation, the solitonic zero-mode Eq.
\ref{soliton} has real coefficients. 
Note that for a given choice of $m_{AF}$, only a
single\footnote{Note that this mode is not electron-hole symmetric}
zero mode exists.
It is also worth to note that time-reversal symmetry $\Theta$ is not a symmetry of the interface. As a result, for the time-reversal counterpart
of the previous system, the zero-mode excitation will be 
$\Theta \Psi^\dagger \Theta^{-1}$,
different from $\Psi^\dagger$. Intuitively, the action of time reversal
symmetry is equivalent to switching between positive or negative
magnetic moments.
In the non-interacting Hamiltonian presented, the zero-energy
mode can be derived analytically, yet
an analogous approach is not available
if we replace the mean-field by a quantum antiferromagnet where the
many-body nature of the system is important. 

Although the many-body problem is challenging, we may connect
with the previous solitonic mode by extending the many-body
Hamiltonian with a staggered field on the Mott side, i.e.  
$H = H_{kin} + H_U + H_{SC} +H_{AF}$. In this way, we introduce
a static moment in addition to the quantum fluctuation. This
model shall again be solved by our computational many-body
scheme.
The schematic result obtained is shown in Fig. \ref{fig4}(a), that
shows that the two time-reversal related solutions found
in the single-particle case, merge in the pure quantum
limit yielding localized in-gap mode.
The previous sketch captures only the single-particle charge excitations reflected in the correlator Eq. \ref{eq:a},
whereas the many-body spectrum will show a continuum of states stemming
from the gapless spinon modes of the quantum antiferromagnet.
The transition from the quantum to the classical regime as the
stagger magnetization is switched on can be
directly observed in the expectation value of the local magnetic moment,
as shown in Fig. \ref{fig4}(b).

We now verify the previous picture by
examining the spectral function Eq. \ref{eq:a} at the interface, 
$n=0$ (Fig.\ref{fig4}(c)) and at $n=10$ inside the Mott region
(Fig.\ref{fig4}(d)). We can observe how the in-gap mode is 
present for $ m_{AF} = 0 $ and gradually transforms into
the zero-energy solitonic mode just described, while
the spectrum within the Mott region remains gapped. 
Thus, the
in-gap spectrum of the many-body system is adiabatically
connected to time-reversal symmetry breaking situation
where low-energy quantum fluctuations are progressively
suppressed upon increasing $ |m_{AF} | $. 

An interesting feature is the splitting of in-gap mode into two branches 
when $ m_{AF} $ is switched on, whereby only one branch evolves
into the solitonic zero-energy mode, while second rises in energy 
and gradually loses weight. 
Moreover, it is also important to note that
depending on the sign of $m_{AF}$,
the low-energy mode will transform either into $\Psi^\dagger$
or into $\Theta \Psi^\dagger \Theta^{-1}$.
In the quantum antiferromagnetic regime, the two modes coexist,
such that there is a two-fold degeneracy for the in-gap mode at $ m_{AF}=0 $, 
whose energy needs not to lie at exactly zero. 

Finally, we highlight two potential platforms to experimentally
realize our proposal,
bulk compounds showing quasi-1D chains and
atomically engineered lattices.
The first direction consists of creating an interface
between a conventional superconductor and a compound
hosting quasi 1D quantum antiferromagnets, such as
CuCl$_2$-2N(C$_5$D$_5$)\cite{PhysRevB.18.3530},
KCuF$_3$\cite{PhysRevLett.70.4003}
and Sr$_2$CuO$_3$.\cite{PhysRevLett.87.247202,PhysRevLett.76.3212}
The second direction consists of exploiting
atomic engineering with atomic scale
microscopy\cite{RevModPhys.91.041001} to create a quantum
antiferromagnet,\cite{RevModPhys.91.041001,Toskovic2016,Loth2012,PhysRevLett.119.227206}
and putting it in contact
with a superconductor.\cite{NadjPerge2014}

\section{Conclusions}
\label{sec:con}

To summarize, we have put forward
a minimal system consisting of
a many-body quantum antiferromagnet
and a conventional s-wave
superconductor that host solitonic
in-gap excitations.
We have unveiled the nature of those states, by showing
that they can be adiabatically connected to solitonic states
between a mean-field antiferromagnet
and a superconductor,
which resembles the Jackiw-Rebbi soliton.
Our results put forward a minimal example in which solitonic
modes appear between a quantum disordered magnet
and a superconductor, providing a stepping
stone towards the study of interfaces between
superconductors and quantum spin liquids.

\section*{Acknowledgments}
M.S. is grateful for the financial support from the Swiss National Science
Foundation (SNSF) through Division II (No.
163186 and 184739).
J.L.L. acknowledges the computational resources provided by the Aalto Science-IT
project.

\section*{Appendix}

\appendix

\renewcommand{\thefigure}{A\arabic{figure}}
\renewcommand{\thesection}{A\arabic{section}}
\renewcommand{\theequation}{A\arabic{equation}}

\newcommand{\clr}{\color{red}}

\begin{figure}[t!]
\centering
    \includegraphics[width=\columnwidth]{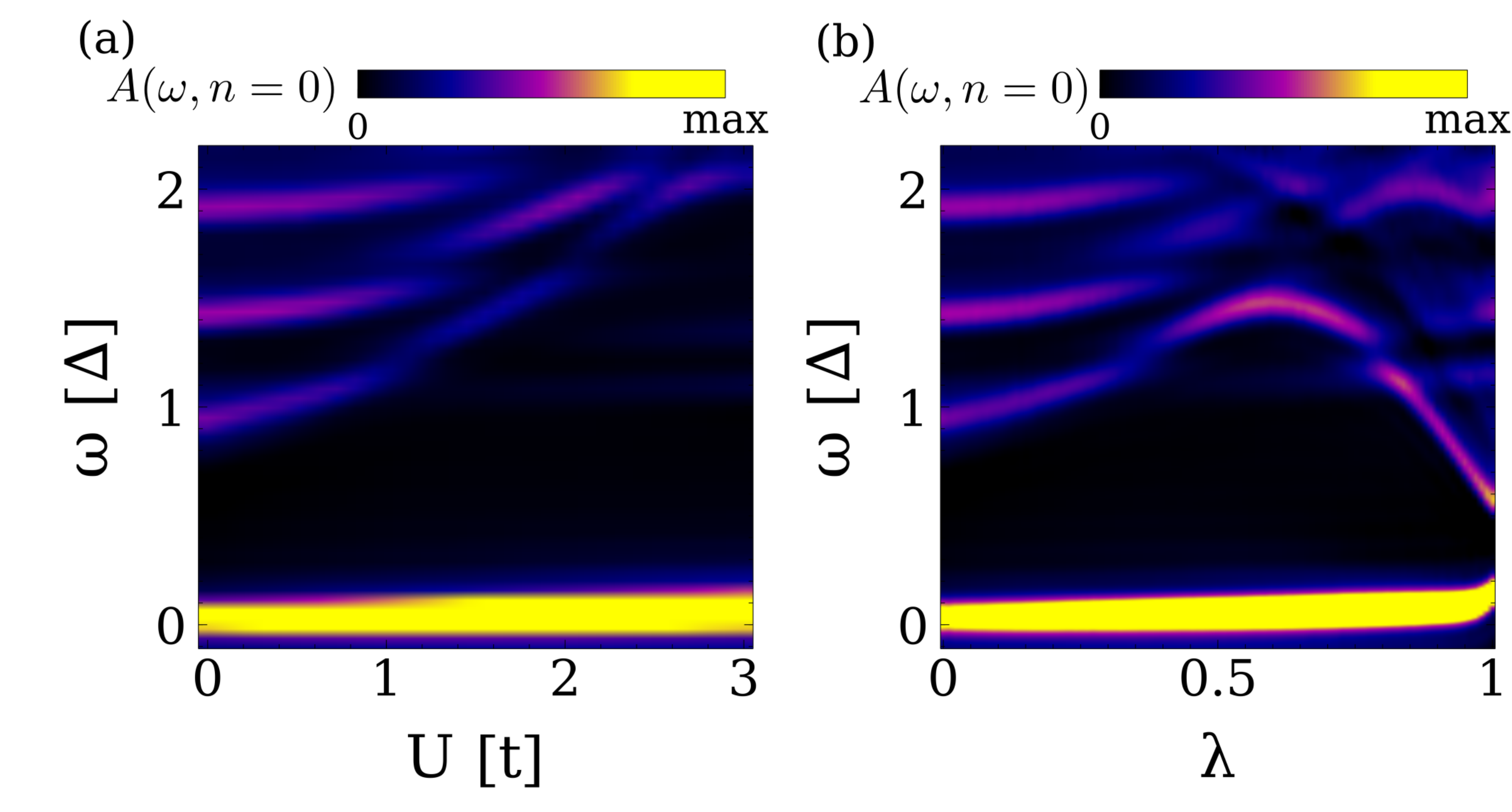}

\caption{
	(a) Evolution of the spectral function at the interface site
	for the Hamiltonian Eq. \ref{eq:hu}, i.e. taking a constant
	$m_{AF}$ and ramping up the value of the Hubbard $U$. Note that
	the whole path has broken time-reversal symmetry.
	Panel (b) shows the interface spectral function for the parametric
	path defined in Eq. \ref{eq:hl}, where $\lambda=0$ denotes
	the analytically solvable limit, and $\lambda=1$ the quantum
	limit (with time reversal symmetry). It is observed that the
	solitonic mode remains robust in both paths, keeping a
	finite bulk gap. We took $\Delta=0.5t$ for (a,b), $U=5t$ for (b)
	and $m_{AF}=0.3t$ for (a,b).
}
\label{fig:inter}
\end{figure}

\section{Adiabatic connection between the mean-field and many-body limit}
In this section we show alternative paths between a free and
interacting limit, complementary to the results of Fig. 4 in the
main text.
We will analyze two cases. First, we connect the
mean-field antiferromagnet and the interacting system,
keeping the stagger magnetization. Second, we connect the mean-field
antiferromagnet directly to the many-body time reversal state.
We elaborate on those two cases below.

First, we show in Fig.
\ref{fig:inter}a the evolution of the interface spectral function
defining a parametric Hamiltonian
\begin{equation}
	H(U) = H_{kin} + H_{SC} + H_{AF} + H_{U}(U)
	\label{eq:hu}
\end{equation}

keeping a fixed m$_{AF}$ and changing $U$.
We observe that the solitonic mode exists in the whole range of this
alternative parametric path. We note that 
time reversal symmetry
remains always broken
due to the presence of a finite $m_{AF}$.

Furthermore, to demonstrate the robustness of the adiabatic connection
used in the main text,
we show an alternative interpolation between the quantum and classical
antiferromagnet. For this purpose, we now define the parametric
Hamiltonian as

\begin{equation}
	H(\lambda) = H_{kin} + H_{SC} + (1-\lambda)H_{AF} + \lambda H_{U}
	\label{eq:hl}
\end{equation}
so that for $\lambda=0$ the Hamiltonian becomes purely non-interacting
(breaking time reversal symmetry), whereas for $\lambda=1$ the
system becomes purely many-body (conserving time reversal symmetry).
As it is observed in Fig. \ref{fig:inter}b, the solitonic
mode exists again in the
whole parametric range, demonstrating its robustness.

It is interesting to note that, since the
interface mode is not of topological origin,
there is not a symmetry protected topological index associated
with it. This is what allows us to connect smoothly the
symmetry broken state, and the time reversal symmetric many-body soliton.
We highlight that along this path, the bulk charge gap of the antiferromagnet
remains open, so that the evolution of solitonic mode can be clearly
followed.

Finally, we note that although the emergence of in-gap states
at interfaces between time-reversal symmetry broken states
and superconductors is a generic feature,\cite{RevModPhys.78.373}
and has been
shown also for antiferromagnetic interfaces.\cite{PhysRevX.5.041042,PhysRevLett.121.037002,PhysRevB.100.125411,PhysRevB.72.184510,Zhen2019}
However, in the
present case we have shown that a robust zero mode appears in the
presence of time-reversal symmetry conservation,
and therefore represents a case dramatically different
from conventional Yu-Shiba-Rusinov states.\cite{RevModPhys.78.373}

\begin{figure}[t!]
\centering
    \includegraphics[width=\columnwidth]{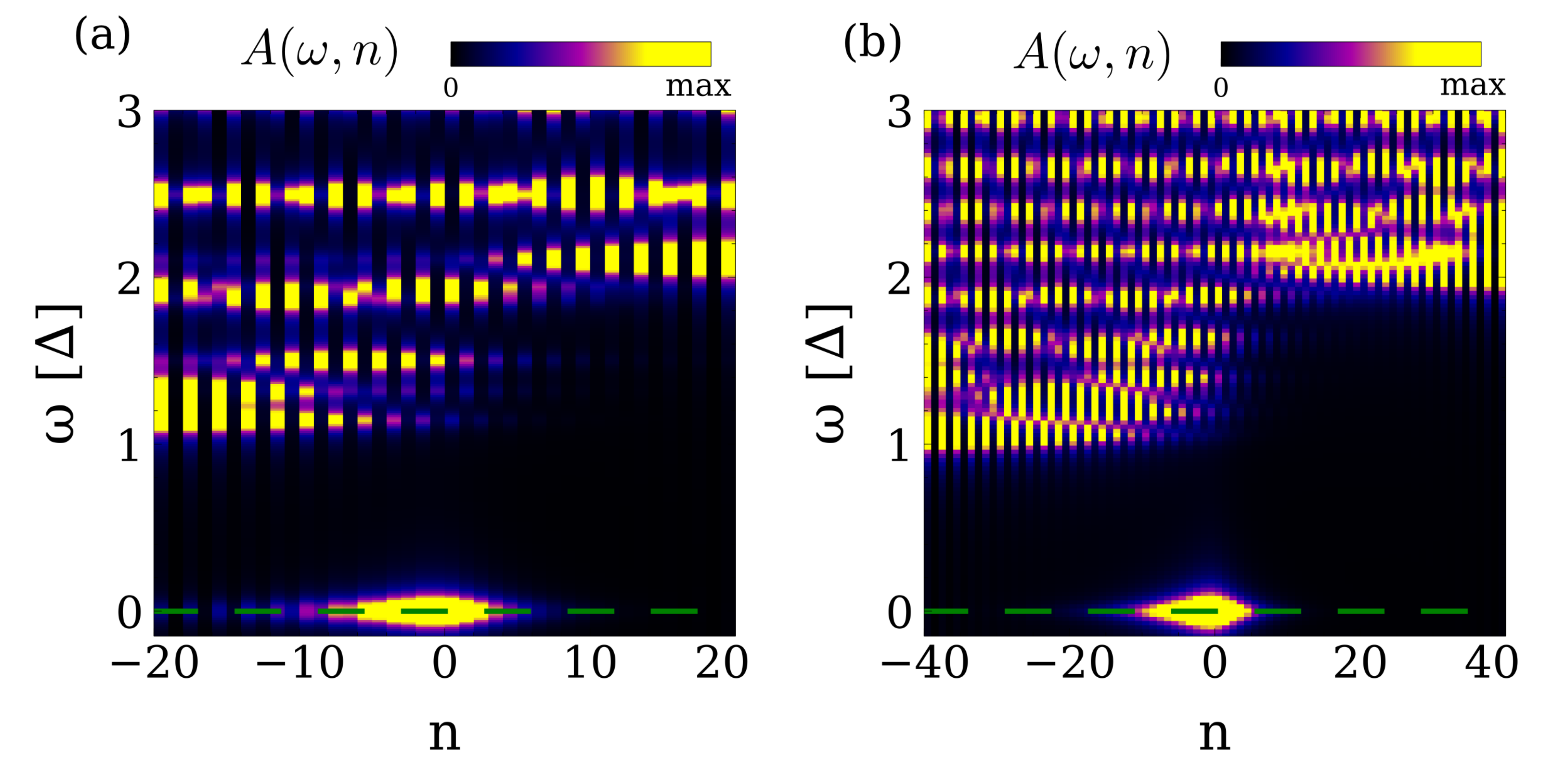}

\caption{Spectral function in the different sites,
for an interface
between a mean-field antiferromagnet and a superconductor,
showing that zero mode is robust with respect
	to the size of the chain, 40 sites for (a) and 80 sites for (b).
	We took $m_{AF}=0.4t$ and $\Delta=0.2t$
}
\label{fig:scalemf}
\end{figure}

\section{Finite size scaling}

In this section we show that the interface solitonic excitation becomes
independent of the length of the chain for large chain size.

We first focus on the effect of different lengths for the
analytically solvable mean-field antiferromagnet.
We first show in Fig. \ref{fig:scalemf} the spectral function in the
non-interacting limit for chains with $L=40$ and $L=80$ sites
(besides the $L=200$ case shown in the main manuscript), highlighting
that the zero mode does not change once chains are sufficiently long.
This exemplifies that the interface mode for the $L=200$ chain used
in our main manuscript is qualitatively analogous to the one
for $L=40$ and $L=80$ of Fig. \ref{fig:scalemf}. We note that this
case is easily
solvable due to the single particle nature of the system.

\begin{figure}[t!]
\centering
    \includegraphics[width=\columnwidth]{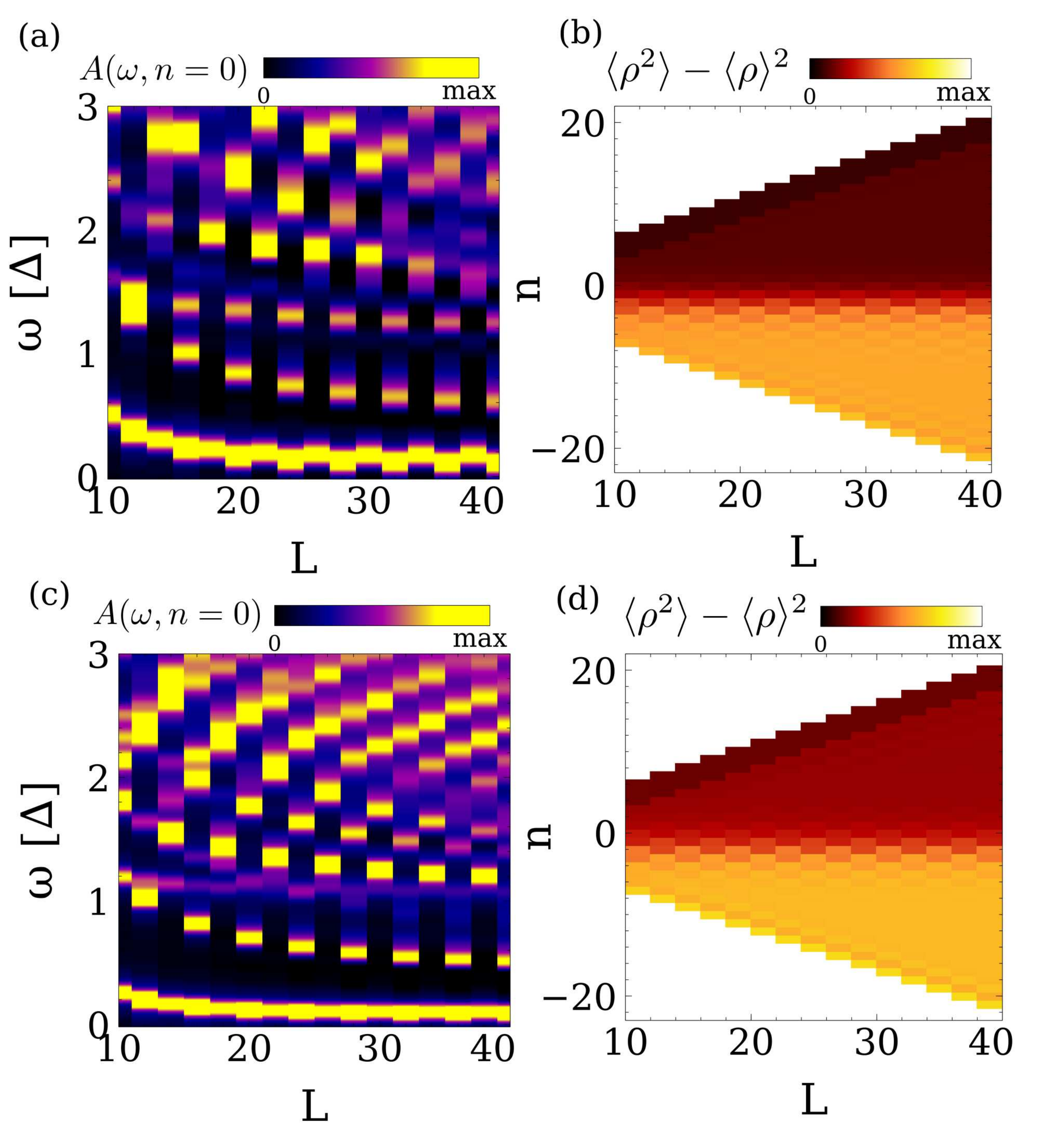}

\caption{ Evolution of the spectral function at the interface
	as a function of the full length of the system $L$ (a,c).
	Evolution of the density fluctuations in real space as a function
	of the chain length $L$ (b,d). Panels
	(a.b) are computed with $U=6t$ and $\Delta=0.4t$, whereas
	panels (c,d) with $U=4t$ and $\Delta=0.6t$. It is observed
	that the solitonic zero mode becomes independent of the
	system size $L$ for large $L$, and
	remains insensitive to the boundary conditions at the interface.
	In contrast, the second in-gap state
	is highly sensitive to the boundary conditions at the interface.
}
\label{fig:qsize}
\end{figure}

We now address the purely many-body quantum limit.
In particular, we have computed how the spectral
function at the interface evolves with the size of the system
as shown in Fig.\ref{fig:qsize}.
As it is observed, the solitonic zero mode remains robust
for different system sizes (Fig.\ref{fig:qsize}a,c). 
In contrast, the second bound mode is sensitive
to the size of the system, that slightly changes
the Hamiltonian at the interface (Fig.\ref{fig:qsize}b,d).
This result illustrates the robustness of the solitonic
mode with respect to the system size,
and justifies once more that
with a $L=40$ chain we reach already asymptotic results.

\section{Spin excitations}

In this section we address the interplay between the solitonic
interface mode and the gapless spinon excitations
of the antiferromagnet.

The solitonic excitation
appears in the charge channel, in which both the superconductor
and antiferromagnet are gaped. The quantum antiferromagnet is gapless
only in the spin channel, where the gapless excitations are spinons.
This suggests that the solitonic mode will be delocalized in the
spin sector, yet localized in the charge sector.
To illustrate this, we have
computed the dynamical response
spin response, defined as

\begin{figure}[t!]
\centering
    \includegraphics[width=0.8\columnwidth]{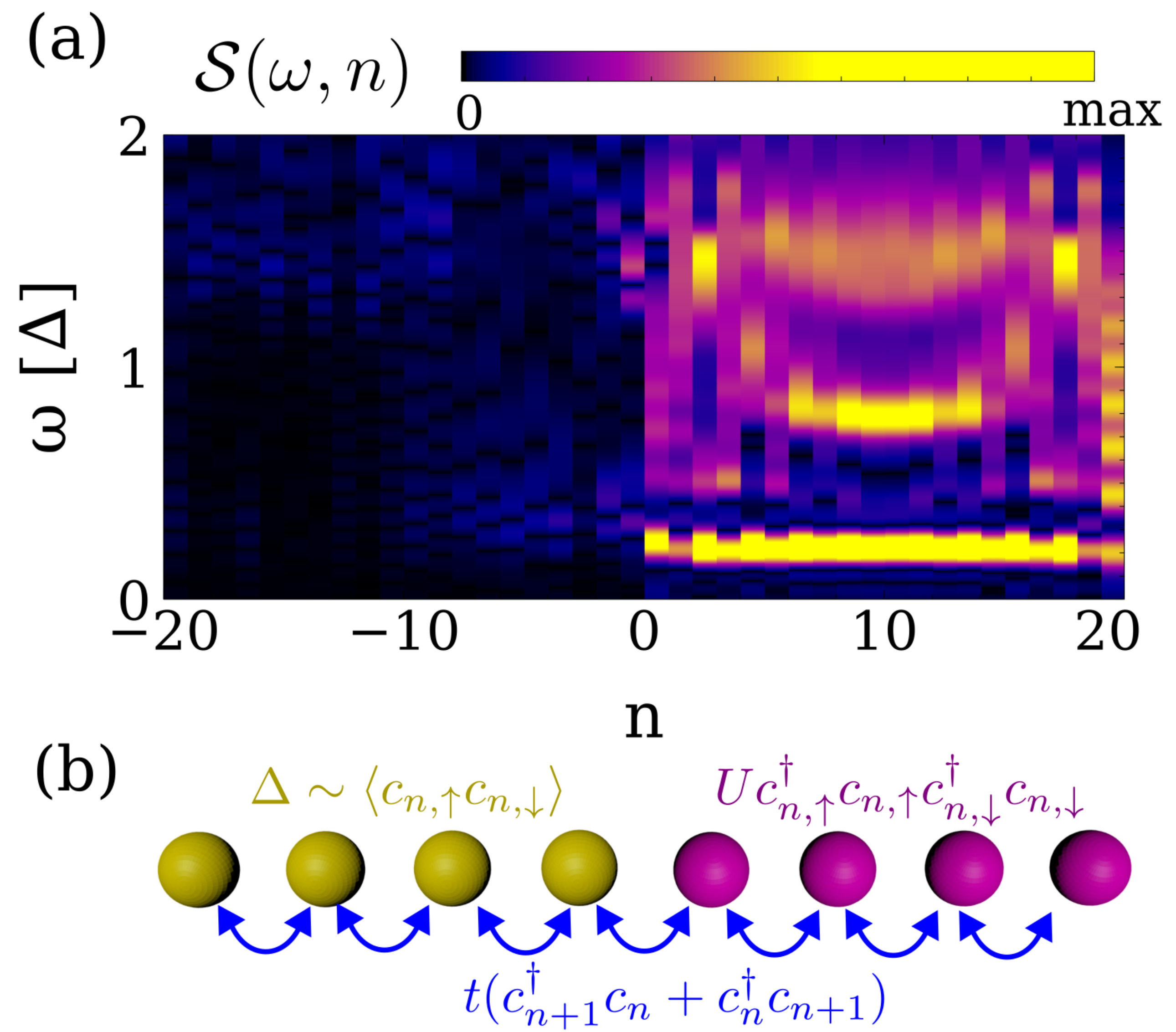}

\caption{
	(a) Dynamical spin response function
	$\mathcal{S} (\omega,n)$ of the superconducting-quantum
	antiferromagnet chain, as defined in Eq. \ref{eq:s}. It is observed
	that the quantum antiferromagnet shows excitations at low energies,
	which are associated with spinons of the Hubbard part. The solitonic
	excitation cannot be easily distinguished in this channel.
	Panel (b) shows a sketch
	of the model.
}
\label{fig:s}
\end{figure}

\begin{equation}
	\mathcal{S} (\omega,n) =
	\langle GS | S_n^z \delta (\omega - H +E_{GS}) S_z^z | GS \rangle
	\label{eq:s}
\end{equation}
show in Fig. \ref{fig:s}. 
The solitonic mode cannot be easily
distinguished in this channel, while
only the gapless low energy modes
of the quantum antiferromagnet
can be observed.
This coexistence suggests than the 
spin sector of the solitonic mode becomes
completely delocalized in the spinon bath, whereas the charge part
remains confined to the interface,
visible in the charge correlator shown in the main text.

\section{Robustness of the soliton mode towards perturbations}
\label{sec:per}

In our main text we have focused in the a minimal model
for the sake of clarity.
We now explicitly show
that the details of the superconductor 
or the existence of additional perturbations
do not matter for the
existance of the solitonic mode. 
These results demonstrate
that the zero mode survives a variety of perturbations
present in a real system, and therefore can be experimentally
observable.

We now elaborate on the different perturbations addressed, which
are summarized below.

\begin{itemize}
	\item Arbitrary doping in the superconductor (Eq. \ref{eq:dop})
	\item Interface scattering (Eq. \ref{eq:ps})
	\item Extended hopping  (Eq. \ref{eq:nnn})
	\item Extended many-body
		interactions (Eq. \ref{eq:v})
	\item Anderson disorder (Eq. \ref{eq:dis})
	\item Selfconsistent treatment of the superconductor
		(Eq. \ref{eq:scf})
\end{itemize}

\begin{figure}[t!]
\centering
    \includegraphics[width=\columnwidth]{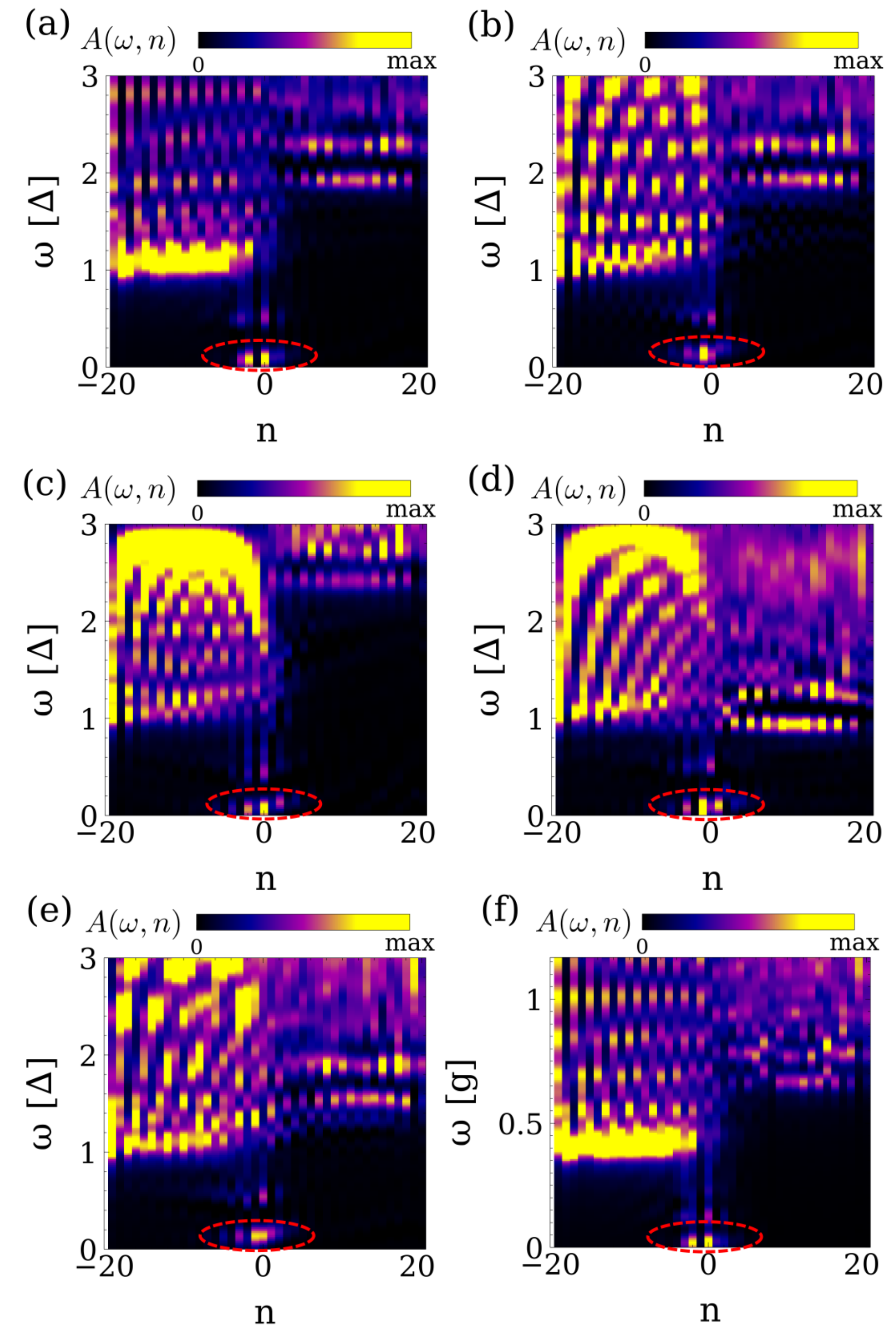}

\caption{Spectral function of the superconductor-quantum
antiferromagnet system with different kinds of perturbations:
	(a) doping in the superconductor Eq. \ref{eq:dop}, 
	(b) interfacial 
	potential scattering
	Eq. \ref{eq:ps}, (c) extended hopping Eq. \ref{eq:nnn},
	(d) extended interactions Eq. \ref{eq:v},
	(e) Anderson disorder Eq. \ref{eq:dis}, and
	(f) selfconsistent superconducting state Eq. \ref{eq:scf}.
	It is observed that in all the instances the solitonic mode
	persists, demonstrating its robustness in
	realistic regimes.
	We took $\Delta = 0.6t$ in (a-e), $U=5t$
	in (a-f) and $N=40$.
}
\label{fig:per}
\end{figure}

In all those instances we have observed the persistence of the
many-body solitonic mode
in our calculations (Fig. \ref{fig:per}). 
We now elaborate on the
results for the different terms considered.

The unperturbed Hamiltonian considered for the system is
$ H = H_{kin} + H_U + H_{SC} $,
with the kinetic term
$
H_{kin} = t\sum_ {n,s} [c^\dagger_{n,s}c_{n+1,s} +
c^\dagger_{n+1,s}c_{n,s}] +
\sum_{n,s} \mu(n)
c^\dagger_{n,s}c_{n,s} \
$, the local interactions of the form
$
H_{U} =
\sum_{n,s} U(n)
c^\dagger_{n,\uparrow}c_{n,\uparrow}
c^\dagger_{n,\downarrow}c_{n,\downarrow} 
$, and the superconducting term of the form
$
H_{SC} =
        \sum_n\Delta(n) [c_{n\uparrow} c_{n\downarrow} +
c^\dagger_{n\downarrow} c^\dagger_{n\uparrow} ] $
as considered in the main manuscript. 
$\Delta(n)$ is defined to be non-zero in the superconductor,
$U(n)$ to be non-zero in the quantum antiferromagnet and $\mu(n)$
is a local onsite energy.
In the following we will add
a variety of perturbations to the previous Hamiltonian,
and show that the zero mode remains present.

First (Fig. \ref{fig:per}a), we consider the case of a superconductor
with an arbitrary doping. For that sake we define a new term
that acts as a chemical potential in the superconducting region.

\begin{equation}
	H_D = D \sum_{i\in SC,s} c^\dagger_{i,s} c_{i,s}
	\label{eq:dop}
\end{equation}
where $i\in SC$ denotes sum over the superconducting part,
and we compute the spectral function for the Hamiltonian
$\bar H = H + H_D$
where we take $D=1.6t$. The result is shown in Fig. \ref{fig:per}a,
and it is clearly observed that the interface zero mode
remains robust. We have verified that the same holds for arbitrary
dopings of the superconductor. This robustness demonstrates that
the existence of the zero mode is not related with
the filling of the superconductor.

Second (Fig. \ref{fig:per}b), we consider the existence of potential
scattering in the interface, as would happen if there is an impurity
at the interface between the antiferromagnet and the
superconductor. The local scattering is implemented in terms
of a local potential at the interface

\begin{equation}
	H_{PS} = w\sum_{s} c^\dagger_{0,s} c_{0,s}
	\label{eq:ps}
\end{equation}
so that the total Hamiltonian is $\bar H = H + H_{PS}$,
and we took $w=0.8t$. As it is observed
in Fig. \ref{fig:per}b the solitonic mode persists in the presence of
potential scattering.

Third (Fig. \ref{fig:per}c), we consider the 
existence of second neighbor coupling in our model
\begin{equation}
	H_{NNN} = t_{NNN}\sum_{n,s} c^\dagger_{n,s} c_{n+2,s} + h.c.
	\label{eq:nnn}
\end{equation}
which breaks the bipartite nature of our model, and generalizes to
realistic realizations
where it is expected a finite second neighbor hopping.
We take $t_{NNN}=0.2t$, and
compute the spectral function for the model $\bar H = H + H_{NNN}$,
whose result is shown in Fig. \ref{fig:per}c. It is observed that the
interface zero mode remains robust, even in the presence of extended
hopping in the model. We have also
verified that the zero mode is also robust
if the second neighbor hopping is included only in the superconductor
or only in the quantum antiferromagnet.

Fourth (Fig. \ref{fig:per}d), 
we consider the effect of nearest neighbor many-body
interactions. In particular, we consider an additional interaction
term of the form

\begin{equation}
	H_{V} = V\sum_{n} 
	\left ( \sum_s c^\dagger_{n,s} c_{n,s} \right )
	\left ( \sum_s c^\dagger_{n+1,s} c_{n+1,s} \right )
	\label{eq:v}
\end{equation}
that acts in the whole system, so that the total Hamiltonian
is $\bar H = H + H_{V}$ and we take $V =0.3t$.
As it is observed in Fig. \ref{fig:per}d the zero mode persists
even in the presence of this additional interaction term.
We have also verified that the zero mode remains
if the interaction is only considered in the superconducting
or quantum antiferromagnetic part.

Fifth, we consider the effect of random Anderson disorder in
the full system as

\begin{equation}
	H_{A} = \sum_{n,s} \delta_n c^\dagger_{n,s} c_{n,s}
	\label{eq:dis}
\end{equation}
where $\delta_n$ is a random number for each site $i$ between
in the interval $[0,0.4t]$. The total Hamiltonian considered
$\bar H = H + H_{A}$, and as shown in Fig. \ref{fig:per}e
it is observed that the zero mode remains present even
in the presence of disorder. We have verified that the zero mode
also remains if disorder is only included in
the superconductor or antiferromagnet.

Finally, we consider the effect of a full self-consistent 
pairing. For this purpose, instead
of imposing a superconducting
pairing $H_{SC}$, we now start with an attractive
interaction in the superconducting region of the
form
$
	H_{g} = -g\sum_{n\in SC,s} 
	c^\dagger_{n,\uparrow} c_{n,\uparrow}
	c^\dagger_{n,\downarrow} c_{n,\downarrow}
$
where $i\in SC$ denotes sum over the superconducting part.
We perform a mean-field decoupling giving rise to
\begin{equation}
	H^{MF}_{g} = -g\sum_{n\in SC,s} 
	\langle c^\dagger_{n,\uparrow} c^\dagger_{n,\downarrow} \rangle
	c_{n,\downarrow} c_{n,\uparrow}
	+ h. c.
	\label{eq:scf}
\end{equation}
so that the total Hamiltonian is
$\bar H = H_{kin} + H_{U} + H_D + H^{MF}_g$. The normal
term of the mean-field decoupling is reabsorbed in $H_{kin}$,
and we take $g = 1.7t$.
The term
$H^{MF}_{g}$ is computed selfconsistently with the tensor network
formalism. We note that this procedure treats the superconductor
at the mean-field level, yielding a selfconsistent
superfluid density, whereas the antiferromagnet is still
treated with the full many-body formalism.
The results are shown in Fig. \ref{fig:per}f, and it is clearly
observed that the solitonic zero mode remains present when
the superconducting term is computed selfconsistently.

\section{Experimental realization}
In this section we present potential
platforms to realize our model experimentally.
Our proposal could be realized
in two different ways, 
with bulk oxides showing quasi-1D chains or
with atomically engineered lattices.
We elaborate on this below.

\begin{figure}[t!]
\centering
    \includegraphics[width=\columnwidth]{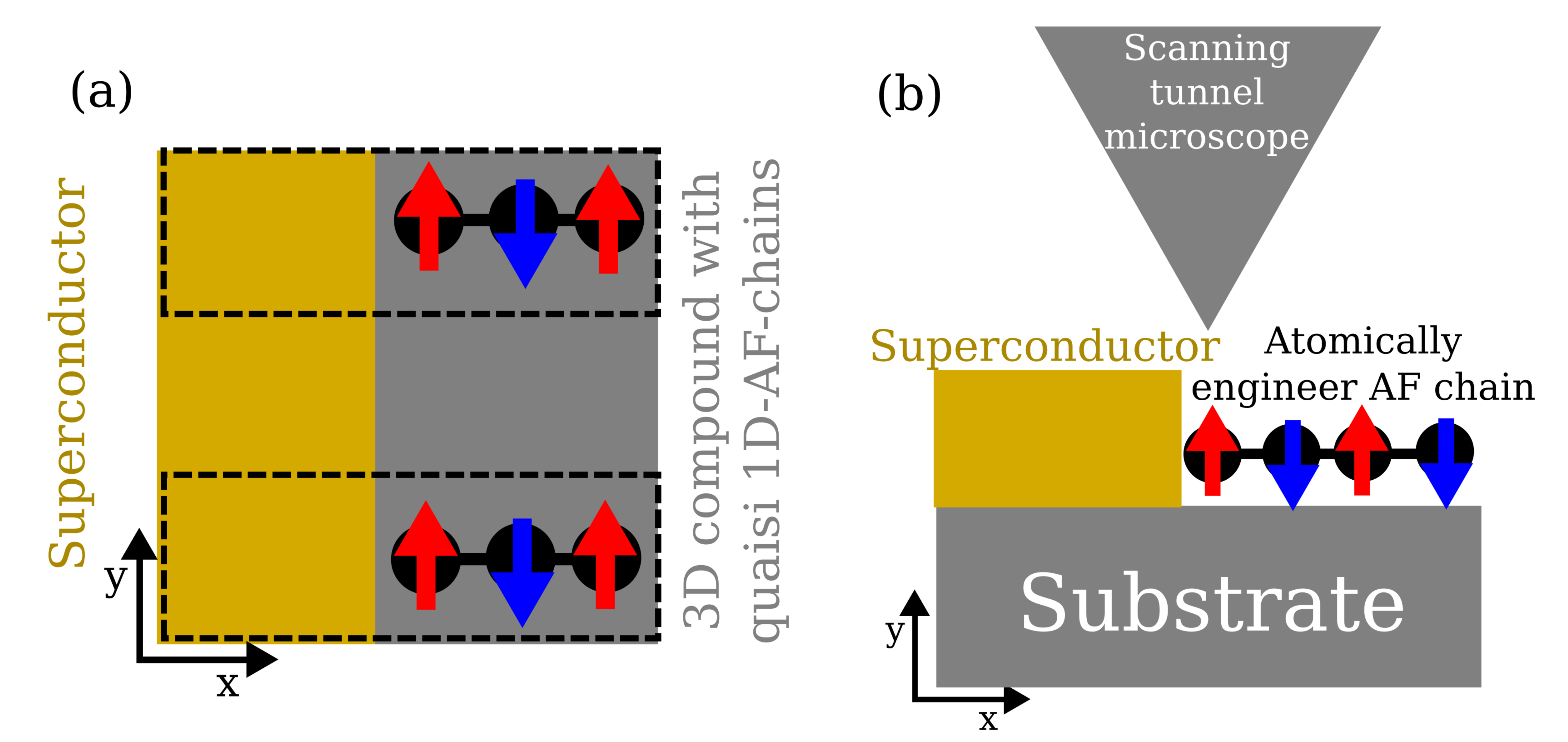}

\caption{
        (a) Sketch of an experimental realization
	of our proposal using a three-dimensional compound
	hosting quasi 1D AF chains.
	Panel (b) shows a sketch of a realization of our proposal
	using atomically engineered spin chains.
}
\label{fig:sketch}
\end{figure}

We first address the proposal based on quasi-1d chains in a three
dimensional compound.
This procedure consist on creating a junction between
a conventional superconductor and a material
hosting nearly decoupled one-dimensional $S=1/2$ 
antiferromagnetic chains, as shown in Fig. \ref{fig:sketch}.
Different
compounds have been extensively studied showing
quasi-1D physics associated
to a strongly interacting Hubbard model,
including CuCl$_2$-2N(C$_5$D$_5$)\cite{PhysRevB.18.3530},
KCuF$_3$\cite{PhysRevLett.70.4003}
and Sr$_2$CuO$_3$.\cite{PhysRevLett.87.247202,PhysRevLett.76.3212}
The interface should be perpendicular to the direction
of the antiferromagnetic chains, as shown in Fig. \ref{fig:sketch}a.
Those compounds have been characterized as to realize an
isotropic Heisenberg model.
Taking an interface of any of those compounds with a conventional
superconductor would lead to a realization of the scenario
proposed in our manuscript.
We
note that although the superconductor is not three
dimensional, the emergence of the zero mode does not depend on the
details of the superconducting part as elaborated in
Section \ref{sec:per}.

We now address the proposal based on 
atomically engineered chains.\cite{RevModPhys.91.041001}
This realization is based on atomic-scale manipulation of individual atoms
using an scanning tunneling microscope (STM), which allows to create atomically
precise structures with specific atoms. These experimental developments
have allowed to realize, at the atomic level, a plethora of paradigmatic
models, including 
one-dimensional quantum critical models,\cite{Toskovic2016}
one-dimensional antiferromagnets,\cite{Loth2012} and
atomic-scale ferromagnets with superconductors,\cite{NadjPerge2014} 
among others.\cite{RevModPhys.91.041001}
The realization with this platform would
require creating a one-dimensional
antiferromagnetic Heisenberg
chain, laterally contacted with a 
superconductor as shown in Fig. \ref{fig:sketch}.
We note that all the ingredients to realize this
structure have been demonstrated,
including 
quantum antiferromagnetism
in $S=1/2$ systems\cite{PhysRevLett.119.227206}.
and superconductivity in
combination with in atomic-scale
engineered chains.\cite{NadjPerge2014}

\bibliographystyle{apsrev4-1}
\bibliography{response}{}

\end{document}